\documentstyle[12pt]{article}
\begin{document}
\title{ Ballistic Four-Terminal Josephson Junction:
\protect\\
Bistable States and Magnetic Flux Transfer}
\author{ A.N.\ Omelyanchouk~${}^{a}$
\footnote{Corresponding author. Email: omelyanchouk@ilt.kharkov.ua}
 \  and  Malek Zareyan~${}^{b}$
\\
{\small {\em ${}^{a}$ B.Verkin Institute for Low Temperature Physics and
Engineering,}}
\\
{\small {\em National Academy of Sciences of Ukraine,}}
{\small {\em 47 Lenin Ave., 310164 Kharkov, Ukraine}}
\\
{\small {\em ${}^{b}$ Institute for Advanced Studies in Basic Sciences,}}
\\
{\small {\em 45195-159, Gava Zang, Zanjan, Iran}}
\\
}
\date{}
\maketitle
\begin{abstract}
The macroscopic quantum interference effects in ballistic Josephson microstructures are investigated. The studied system consists of four bulk superconductors (terminals) which are weakly coupled through the mesoscopic rectangular normal metal (two dimensional electron gas). We show that nonlocal electrodynamics of ballistic systems leads to specific current-phase relations for the mesoscopic multiterminal Josephson junction. The nonlocal coupling of supercurrents produces the "dragging" effect for phases of the superconducting order parameter in different terminals. The composite Josephson junction, based on this effect, exhibits the two -level system behaviour controlled by the external magnetic flux. The effect of magnetic flux transfer in a system of nonlocally coupled superconducting rings is studied.
\end{abstract}
\newpage

\section{Introduction}
Recently there has been considerable interest in ballistic mesoscopic S-2DEG-S Josephson junctions (2DEG - two dimensional electron gas) . Heida et.al. \cite{He} investigating this new class of fully phase coherent Josephson junctions of comparable width and length, have demonstrated the nonlocal phase dependence of mesoscopic supercurrents. Nonlocality leads to the anomalous critical current dependence on the magnetic flux in such mesoscopic junctions [1-3]. It was noted in \cite{OOV95} that the four-terminal generalization of S-2DEG-S structures (namely, considering of the case when two-dimensional normal layer is connected with four terminals instead of two ones as in Refs.[1-3]) will allow the studying of the specific multichannel macroscopic quantum interference effects in mesoscopic multiterminals.  For review of experimental and theoretical investigation of the conventional multiterminal Josephson junctions see \cite{OOr} and references there. The microscopic theory of coherent current states in mesoscopic ballistic 4-terminal Josephson junction was offered in \cite{ZO}. In present paper we study the macroscopic manifestations of the nonlocal coupling of supercurrents inside the mesoscopic multiterminal weak link. We show, that 4-terminal Josephson junction with two terminals incorporated in superconducting rings, presents the macroscopic bistable system in which the energy levels and the potential barrier height are regulated by the magnetic flux $\Phi$ applied through the ring. Such composite S-N($\Phi$)-S Josephson junction between conventional superconductors behaves similar to so called "$\pi$-contact" (see the recent paper \cite{Zagd} on tunable qubit, based on "$\pi$-contact" between d-wave superconductors). In addition, we introduce the effect of magnetic flux transfer in system of two superconducting rings weakly coupled through the mesoscopic 4-terminal junction.

\section{System description. Basic equations}
The mesoscopic four-terminal Josephson junction we study is sketched in Fig. 1a.
It consists of four bulk superconductors (S) which are weakly coupled through a ballistic rectangular  normal (N) layer (two dimensional electron gas ) of length $L$ and width $W$. Contact sizes $L$, $W$ are suppose to be much larger than Fermi wave length ${\lambda}_F$. In the stationary state, the ballistic motion of the electrons through the normal junction region ($-L/2<x<L/2$ ,  $-W/2<y<W/2$)  is described by the quasi-classical Eilenberger equations \cite{Eil}:

$$
{\bf v}_F\frac \partial {\partial {\bf r}}\hat G+[\omega \hat \tau _3+\hat \Delta , \hat G]=0,
$$
\begin{equation}
\hat G=\left(
\matrix { g&f\cr
         f^{\dagger}&- g\cr}
         \right), \ \ \
\hat \Delta=\left(
\matrix {0&\Delta\cr
         \Delta^{*}&0\cr}
         \right).
\end{equation}
\par
The electrons go through quasi-classical trajectories which are straight lines along the electron velocity on the Fermi surface, ${\bf v}_F$.
The trajectory which is coming from the $i$th bank  and going to the $j$th bank  ($i\rightarrow j$ trajectory) is shown in Fig. 1b. In the point ${\vec \rho}(x,y)$ , the interference between the trajectories connecting different superconducting banks ({\it e.g.}  $i\rightarrow j$ and $i^{\prime}\rightarrow j^{\prime}$, in Fig. 1b) takes place.  The state of $i$th S-terminal is defined by the complex pair potential
$\Delta=\Delta_0e^{i\varphi _i}$\ $(i=1,...,4)$ ,{\it i.e.} by the phase $\varphi_i$ in the $i$th  bulk superconductor.\par
The solution of Eilenberger equations (1) inside N-layer can be easily obtained by integrating over the electronic trajectories in the same way as for a ballistic point contact \cite{KO}. In Ref. \cite{ZO}, the anomalous $f_{\omega} ( {\bf v}_F,{\vec \rho})$
and normal  $g_{\omega} ( {\bf v}_F,{\vec \rho})$  Green's functions were calculated for arbitrary temperatures $0<T<T_c$ and arbitrary sizes $L,W$ (in the scale $\xi_0\sim\hbar v_F/\Delta_0$) . For the current density ${\bf j}({\vec \rho})$ and self-consistent off-diagonal potential $\Delta$, we have the expressions:

\begin{equation}
{\bf j}({\vec \rho})=-4\pi ieN(0)T\sum\limits_{\omega >0}\left\langle {\bf v}_Fg_\omega
\right\rangle,
\end{equation}
\begin{equation}
\Delta =\lambda 2\pi T\sum\limits_{\omega >0}\left\langle f_\omega
\right\rangle .
\end{equation}
Here  $\omega $ is Matsubara frequency, $N(0)= m/(2\pi),$ $\left\langle ...\right\rangle $ is the averaging over directions of 2D
vector ${\bf v}_F$, $\lambda $ is the constant of electron-phonon coupling.
Inside the normal layer $\lambda=0$ and hence $\Delta=0$. However, the anomalous average,    $( {\hat \psi}{\hat \psi})$ has the finite value and determines the induced order parameter, $ \Psi ({\vec \rho})$, as

\begin{equation}
\Psi ({\vec \rho}) =2\pi T\sum\limits_{\omega >0}\left\langle f_\omega
\right\rangle .
\end{equation}

\section{Current-phase relations }
We consider the case of small junction ($L,W<<\xi_0$) when the effects of nonlocality are most pronounced. Qualitatively,  it describes the situation of $L,W\simeq\xi_0$. For general consideration see \cite{ZO}.
The total current flowing into the $i$th terminal, $I_{i}$, depends on the phases $\varphi _j$ ($j=1,...,4$) in all the banks:

\begin{equation}
I_{i}= \frac{\pi \Delta_0 }{e} \sum_{j=1} ^{4} \gamma_{ij}
\sin{(\frac{\varphi_{i}-\varphi_{j}}{2})}\tanh{[\frac{\Delta_0
\cos{(\frac{\varphi_{i}-\varphi_{j}}{2})}}{2T}]}.
\end{equation}

In the case of two-terminal junction we have from (5) the Josephson current of a clean point contact\cite{KO}. The geometrical factors, $\gamma_{ij}$, present the coupling coefficients between the $i$th and $j$th terminals and are functions of the length to width ratio $k=L/W$ as
$$
\gamma_{13}=\frac{e^2p_Fd}{\pi^2}(1-\frac{1}{\sqrt{1+k^2}}),\ \ \
 \gamma_{24}=\frac{e^2p_Fd}{\pi^2}(1-\frac{k}{\sqrt{1+k^2}})
$$
\begin{equation}
\gamma_{12}=\gamma_{14}=\gamma_{23}=\gamma_{34}=
\frac{e^2p_Fd}{2\pi^2}(\frac{1+k}{\sqrt{1+k^2}}-1),\ \ \                  
 \gamma_{ij}=\gamma_{ji}.
\end{equation}
where $d=\sqrt{L^2+W^2}$ and the numeration of the banks is shown in Fig. 1b.
In eq. (5), the positive sign of $I_i$ corresponds to the direction of the current
from the normal layer to the $i$th bank. Note that $\sum_{i=1} ^{4} I_{i} =0$.\\
In the following we consider, for simplicity, the symmetric junction (square N-layer), $k=1$, and the case of temperatures $T$ near the critical temperature $T_c$. We emphasize, that the studied here macroscopic interference phenomena take place at low temperatures, {\it e. g.} at $T=0$, as well as near $T_c$. The  difference is in the  values of the critical currents and in the replacing of the sinusoidal current-phase dependence $\sin{\varphi}$ by
$\sin{(\varphi/2)}~ sign{[\cos{(\varphi/2)}]}$.

Near $T_c$,  and for $k=1$, eq. (5) takes the form

\begin{equation}
I_{i}= \frac{\pi{\Delta^2 _0}(T)}{4e T_c}\frac{e^2p_FL}{\pi^2}\frac{\sqrt{2}-1}{\sqrt{2}} \sum_{j=1} ^{4}
{\tilde \gamma_{ij}}\sin{(\varphi_{i}-\varphi_{j})},
\end{equation}
with
\begin{equation}
{\tilde \gamma_{13}}={\tilde \gamma_{24}}=\sqrt{2},\ \ \ 
 {\tilde \gamma_{12}}={\tilde \gamma_{14}}={\tilde \gamma_{23}}={\tilde \gamma_{34}}=1.
\end{equation}
The corresponding Josephson coupling energy of the 4-terminal junction, $E_J$, (such that $I_i=(2e/\hbar)\partial E_{J}/\partial \varphi_i$) has the form

\begin{equation}
E_J({\varphi_{i}})=\frac{\hbar}{2e}\frac{\pi{\Delta_0(T)}^2}{4eT_c}
\frac{e^2p_FL}{\pi^2}\frac{\sqrt{2}-1}{\sqrt{2}}
\sum_{j<k} {\tilde \gamma_{jk}}  [1-\cos{(\varphi_{j}-\varphi_{k})}].
\end{equation}
In the case of conventional 4-terminal Josephson junction \cite{OOV95}, which presents the system of crossed dirty superconducting microbridges between bulk superconductors, the current-phase relations have the form similar to (7). The essential difference consists in the structure of the coupling coefficients $ \gamma_{ij}$. In the mesoscopic ballistic junction these form factors are not factorized (namely, $\gamma_{12}\gamma_{34}\neq  \gamma_{13}\gamma_{24}$), in spite of the conventional case, where $ \gamma_{ij}\sim (1/R_i)(1/R_j) $; $R_i$ are the normal resistances of the microbridges.\par
The current-phase relations (7) determine the behaviour of the system in the presence of the transport currents and (or) the diamagnetic currents induced by
the magnetic fluxes through closed superconducting rings. 
We will distinguish two types of the circuit implication of  the mesoscopic 4-terminal junction (Fig. 2). The first one, "crossed" or "transverse" implication is shown in Fig. 2a. It closely corresponds to the case of conventional 4-terminal junction. Thus, the theory developed in  \cite{OOr} is applicable to this case too. In the "parallel" (Fig. 2b) case, the new specific features in the behaviour of the 4-terminal Josephson junction appear. They are  results of the nonlocal coupling of currents inside the N-layer, which produces the peculiar effect of "dragging" of  the phase difference between one pair of terminals by the phase difference between another pair of terminals.
We will concentrate here on the "dragging" case (Fig. 2b).\par
By introducing the phase differences
$$
\theta={\varphi}_{2}-{\varphi}_{1}, \ \ \
 \varphi={\varphi}_{3}-{\varphi}_{4}, \ \ \
\chi=\frac{{\varphi}_{1}+{\varphi}_{2}}{2}-\frac{{\varphi}_{3}+{\varphi}_{4}}{2},
$$
and taking into account the relations

\begin{equation}
I=I_2=-I_1, \ \ \ J=I_3=-I_4,
\end{equation}
we obtain from (7)

\begin{equation}
I=\sin{\theta}+[(\sqrt{2}+1)\sin{\frac{\theta}{2}}\cos{\frac{\varphi}{2}}+ 
(\sqrt{2}-1)\cos{\frac{\theta}{2}}\sin{\frac{\varphi}{2}}]\cos{\chi},
\end{equation}

\begin{equation}
J=\sin{\varphi}+[(\sqrt{2}+1)\sin{\frac{\varphi}{2}}\cos{\frac{\theta}{2}}+
(\sqrt{2}-1)\cos{\frac{\varphi}{2}}\sin{\frac{\theta}{2}}]\cos{\chi}.
\end{equation}
All the currents are normalized by
$$
I_0=\frac{\pi{\Delta_0(T)}^2}{4eT_c}\frac{e^2p_FL}{{\pi}^2}\frac{(\sqrt{2}-1)}{\sqrt{2}}.
$$
From the current conservation, it follows that the phase $\chi$ in eqs. (11) and (12) can takes only two value of $0$ or $\pi$.
The dimensionless Josephson coupling energy $E_J$ (in units $\hbar/(2e)I_0$) in terms of $\varphi$, $\theta$ and $\chi$ takes the form

\begin{equation}
E_J=-\cos{\theta}-\cos{\varphi}-[(\sqrt{2}+1)\cos{\frac{\varphi}{2}}\cos{\frac{\theta}{2}}-
 (\sqrt{2}-1)\sin{\frac{\varphi}{2}}\sin{\frac{\theta}{2}}]\cos{\chi}.
\end{equation}
The minimization of $E_J$ with respect to $\chi$ actually gives that phase $\chi$ takes the value $0$ or $\pi$, depending on the equilibrium values of $\theta$ and $\varphi$:

\begin{equation}
\cos{\chi}=sign{[(\sqrt{2}+1)\cos{\frac{\varphi}{2}}\cos{\frac{\theta}{2}}-(\sqrt{2}-1)\sin{\frac{\varphi}{2}}\sin{\frac{\theta}{2}}]}.
\end{equation}
The coupled system of equations (11), (12) with condition (14) for $\chi$ determines the current-phase relations $I(\theta;\varphi)$ and $J(\varphi;\theta)$. We note that in the "crossed" implication case (as well as in the conventional 4-terminal Josephson junction) the coupling terms in eqs. (11)- (14), which contain the coefficient $(\sqrt{2}-1)$, are absent. The appearance of this additional coupling leads to  new effects.

\section{Bistable states in composite S-N($\Phi$)-S junction}
Let us consider the case when the terminals 3 and 4 are incorporated in a superconducting ring. When the external flux is applied, the phase $\varphi$ is controlled by the magnetic flux $\Phi$, which is embraced by the ring:  $\varphi =2\pi \Phi/\Phi_0$ ($\Phi_0=h/2e$).
The configuration shown in Fig.3 presents the Josephson weak link (the region enclosed by the dashed rectangular) between two superconductors. This weak link has the current-phase relation $I(\theta)$, which depends on the internal degrees of freedom (trapped flux $\Phi$) of such composite S-N($\Phi$)-S junction. The nonlocal coupling in the N-layer leads to the nontrivial behaviour of the $I(\theta )$ as function of $\Phi$. It follows from (11) and (12) that for $\Phi \neq 0$, due to the term proportional to $(\sqrt{2}-1)$, zero value of the current $I$ corresponds to nonvanishing value of the phase $\theta$.
 In Fig. 4 the function $I(\theta)$ is shown for two values of flux,  $\Phi =0$ (dashed curve) and $\Phi =\Phi_0/2$ (solid curve).
The phase dependence $I(\theta)$ is splitted near $\theta=0$ at $\Phi\neq 0$. If $\Phi =\Phi_0/2$  and $I=0$ the phase difference is not zero but equals to $\theta=\pm \theta_0=\pm \arcsin{(\frac{\sqrt{2}-1}{\sqrt{2}+1})}$. Both values  $\pm \theta_0$ correspond to the equal minima of potential $E_J(\theta|\Phi=\Phi_0/2)$ as shown in Fig.5. With going of $\Phi $ apart from $\Phi_0/2$ the potential profile is changed as shown in Fig.5, and one of the minima disappears.
Thus, the S-N($\Phi$)-S junction presents the two level system  with energy levels which can be regulated by the magnetic flux $\Phi$. At   $\Phi =\Phi_0/2$ we have the bistable system with equal energy values of the levels.  The picture resembles  the so called "$\pi$-contacts", {\it i.e.} the conventional Josephson junction with magnetic impurities in barrier or contact  between d-wave superconductors.

\section{Magnetic flux transfer}
The appearance of  nonzero phase difference $\theta$ at zero value of current $I$ in
 S-N($\Phi$)-S system leads to the effect of magnetic flux transfer, which is a macroscopic manifestation of the nonlocal coupling in the mesoscopic 4-terminal Josephson junction.\par
Consider the case when both pairs of terminals 1-2 and 3-4 are short circuited by superconducting rings (  S-N($\Phi$)-S junction incorporated in the ring). The currents $I$ and $J$ in eqs. (11), (12) are now related to the corresponding magnetic fluxes through the rings by

\begin{equation}
I=\frac{\Theta^e-\Theta}{L_I},
\ \ \ \ \ \
J=\frac{\Phi^e-\Phi}{L_J},
\end{equation}
where $L_I$, $L_J$ are self-inductances of the rings, 
$\Theta^e$ and $\Phi^e$ are the external magnetic fluxes applied to the rings and
$\Theta$ and $\Phi$ are the  resulting embraced fluxes. The phases $\theta$ and $\varphi$ are related to the magnetic fluxes by
$\theta=2\pi \Theta/\Phi _0$,  $\varphi=2\pi\Phi /\Phi_0$.

From eqs. (11), (12) and (15) we obtain the coupled equations for  $ \Theta$ and  $\Phi$

\begin{equation}
\Theta={\Theta}^e-{{\cal
L}_I}\{\sin{\Theta}
+[(\sqrt{2}+1)\sin{\frac{\Theta}{2}}\cos{\frac{\Phi}{2}}+ 
(\sqrt{2}-1)\cos{\frac{\Theta}{2}}\sin{\frac{\Phi}{2}}]\cos{\chi}\}, 
\end{equation}
\begin{equation}
\Phi ={\Phi}^e-{{\cal
L}_J}\{\sin{\Phi}
+[(\sqrt{2}+1)\sin{\frac{\Phi}{2}}\cos{\frac{\Theta}{2}}+ 
(\sqrt{2}-1)\cos{\frac{\Phi}{2}}\sin{\frac{\Theta}{2}}]\cos{\chi}\},
\end{equation}
where we have introduced dimensionless self-inductances ${\cal L}_{I,J}=(2e/\hbar)I_0L_{I,J}$ and magnetic fluxes are measured in units $h/2e$.
It follows from these equations that, when the flux ${\Phi}^e$ is applied to the $J$-ring,  even at ${\Theta}^e=0$ the magnetic flux in  $I$-ring $\Theta$ is not zero. Suppose for simplicity that ${\cal L}_{J}<<1$, so that
${\Phi}\simeq \Phi^e$. For the transferred flux ${\Theta}$ (at ${\Theta}^e=0$) we obtain the equation:

$$
\Theta =-{{\cal
L}_I}\{\sin{\Theta}
+[(\sqrt{2}+1)\sin{\frac{\Theta}{2}}\cos{\frac{\Phi^e}{2}}+
$$
\begin{equation}
(\sqrt{2}-1)\cos{\frac{\Theta}{2}}\sin{\frac{\Phi^e}{2}}]~sign[(\sqrt{2}+1) 
\cos{\frac{\Phi^e}{2}}\cos{\frac{\Theta}{2}}-(\sqrt{2}-1)\sin{\frac{\Phi^e}{2}}\sin{\frac{\Theta}{2}}] \}. 
\end{equation}
In the limit ${\cal L}_{I}\rightarrow 0$, we obtain from (17) for transferred flux
\begin{equation}
\Theta =-{{\cal
L}_I}(\sqrt{2}-1)\sin{\frac{\Phi^e}{2}}~sign{(\cos{\frac{\Phi^e}{2}})}.
\end{equation}
The numerical solutions of eq. (18)  for different values of ${\cal L}_{I}$ are presented in Fig. 6.
The transferred flux $\Theta$ is a periodical function of applied flux $\Phi^e$,
$\Theta(\Phi^e+2n\pi)=\Theta(\Phi^e)$  where $n$ is an integer number.
For finite values of ${\cal L}_{I}$, the small hysteresis region exists in the dependence $\Theta (\Phi^e)$ near $\Phi^e=\Phi_0/2$ (see Fig.6).

\section{Summary}

The multiterminal Josephson junction presents the system in which weak coupling takes place between several massive superconducting banks (terminals). The general phenomenological theory of Josephson multiterminals \cite{OOr} shows the variety of macroscopic quantum interference phenomena in such systems. In present paper, we have introduced and studied a new kind of Josephson multiterminals, the mesoscopic ballistic multiterminal Josephson junction. In the studied system the nonlocal coupling is established through a mesoscopic two dimensional normal metal layer. Within the microscopic theory, the current-phase relations for mesoscopic multiterminal junction have been calculated. They take into account the nonlocal coupling of supercurrents inside the ballistic weak link. This current-phase relations describe specific (as compared to the conventional multiterminals \cite{ OOr}) behaviour of the system in the presence of the transport currents and (or) the diamagnetic currents induced by the magnetic fluxes through the closed superconducting rings. We showed that nonlocality produces the dragging of the phase difference between one pair of terminals by the phase difference between another pair of terminals.\par
We have studied the macroscopic manifestations of this dragging in two situations. The first one is, what we call S-N($\Phi$)-S Josephson junction (Fig. 3). It presents the S-N-S contact with a N-layer which properties  are regulated by the external magnetic flux applied through the ring as shown in Fig. 3. The internal degree of freedom (magnetic flux $\Phi$ embraced in the ring) makes this macroscopic system similar to the $\pi$-contacts, {\it i. e. } to the contacts having non-zero ground state phase difference at zero value of the current. We obtained that our " macroscopic $\pi$-contact" presents the bistable system in which the energy levels and potential barrier are regulated by the magnetic flux $\Phi$. At $\Phi$ equals to 1/2 of magnetic flux quantum $\Phi_0$, we have doubly degenerate ground state, just as in the case of $\pi$-contact between d-wave superconductors (see \cite{Zagd}). \par
Secondly, we consider a system of two superconducting rings weakly coupled through the mesoscopic 4-terminal junction. Such system was studied in \cite{OOV95} for the case of conventional 4-terminal Josephson junction. Now (in addition to the results of Ref.  \cite{OOV95}), in nonlocal mesoscopic case the effect of magnetic flux transfer is obtained. It consists of appearance of magnetic flux in one of the rings when the magnetic flux is applied through the another ring. The amplitude of effect is calculated for different values of the ring's selfinductance.\par
In conclusion, it is interesting to realize the 4-terminal SQUID device \cite{OOr,Velt} based on mesoscopic 4-terminal junction, which will expose the studied here effects of nonlocality.



{\large Figure captions}\\

Fig. 1. (a) The mesoscopic ballistic 4-terminal Josephson junction. Four bulk superconductors (S) are weakly coupled through a rectangular normal (N) layer of length $L$ and width $W$.
b) The electron quasi-classical trajectories inside the normal region ($-L/2<x<L/2,-W/2<y<W/2$) are straight lines along the electron velocity on the Fermi surface ${\bf v}_F$. The trajectory $i\rightarrow j$ connects the $i$th and $j$th sides.\\

Fig. 2. Two types of implication of mesoscopic 4-terminal Josephson junction in the external circuits.
(a) Crossed 
(b) Parallel.\\

Fig. 3. The S-N($\Phi$)-S Josephson junction. Two superconductors (S) are weakly coupled through the composite weak link (dashed rectangular) which consists of the N-layer and a coupled superconducting ring with the embraced magnetic flux $\Phi$.\\

Fig. 4. The current $I(\theta)$ through the S-N($\Phi$)-S Josephson junction is shown for two values of $\Phi$: $\Phi=0$ (dashed curve) and $\Phi=\Phi_0/2$ (solid curve).
For $\Phi=\Phi_0/2$, $I(\theta)$ is splitted near the $\theta=0$.\\

Fig. 5. The Josephson coupling energy for a S-N($\Phi$)-S junction for $\Phi=\Phi_0/2$ (solid curve) has two equals minima. For $\Phi>\Phi_0/2$ or $\Phi<\Phi_0/2$ (dashed curves) the potential profile is changed and one of the minima disappears.\\

Fig. 6.  The dependence of the transferred flux to the $I$-ring, $\Theta$, on the flux in the $J$-ring, $\Phi$,
for different values of selfinductance  ${\cal L}_{I}$:
"-----" -  ${\cal L}_{I}=1$,
"$\circ \circ \circ $" - ${\cal L}_{I}=0.5$,
"$ \Box \Box \Box$" - ${\cal L}_{I}=0.1$. The dashed rectangular shows the smal hysteress region near $\Phi=\pi$.

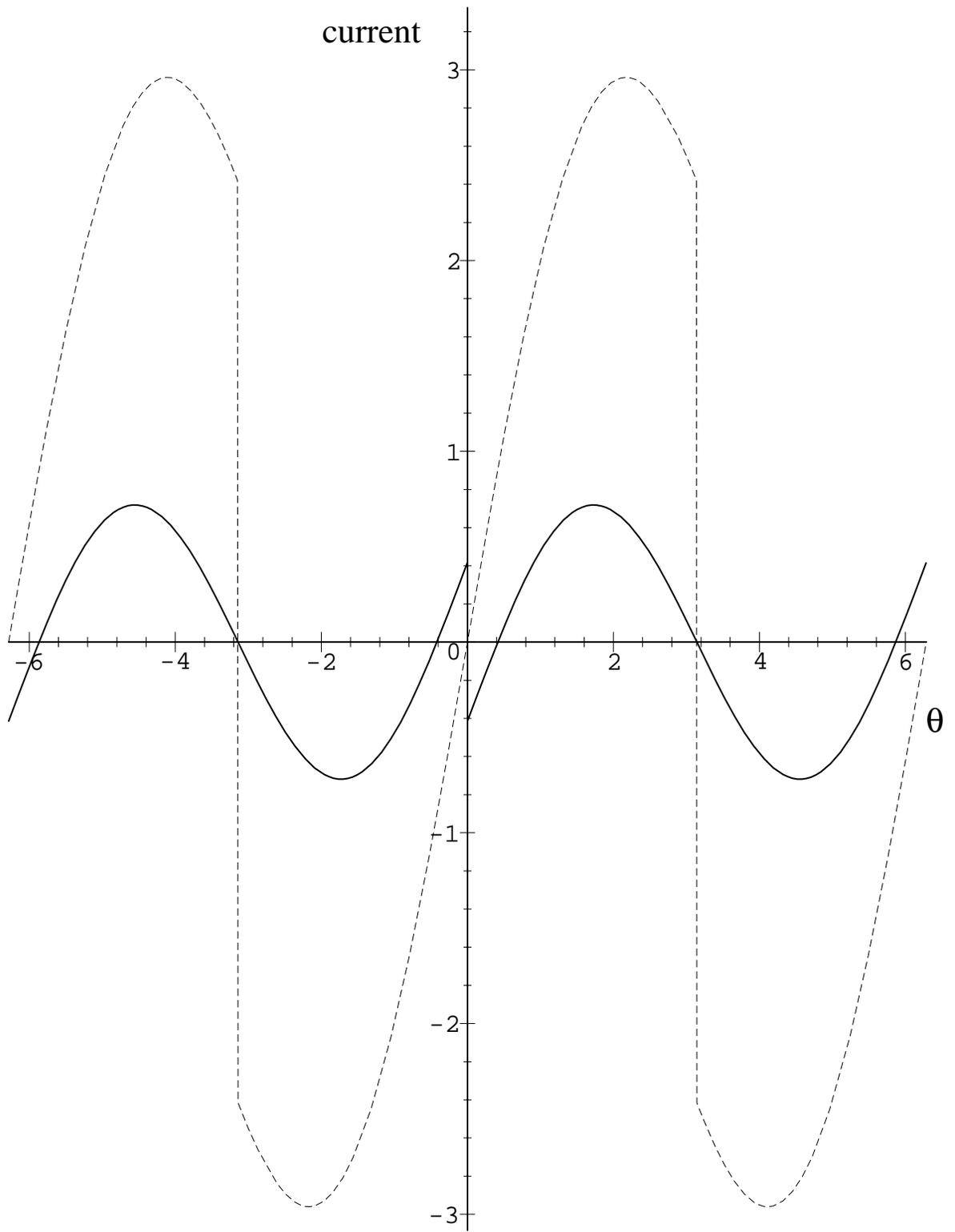
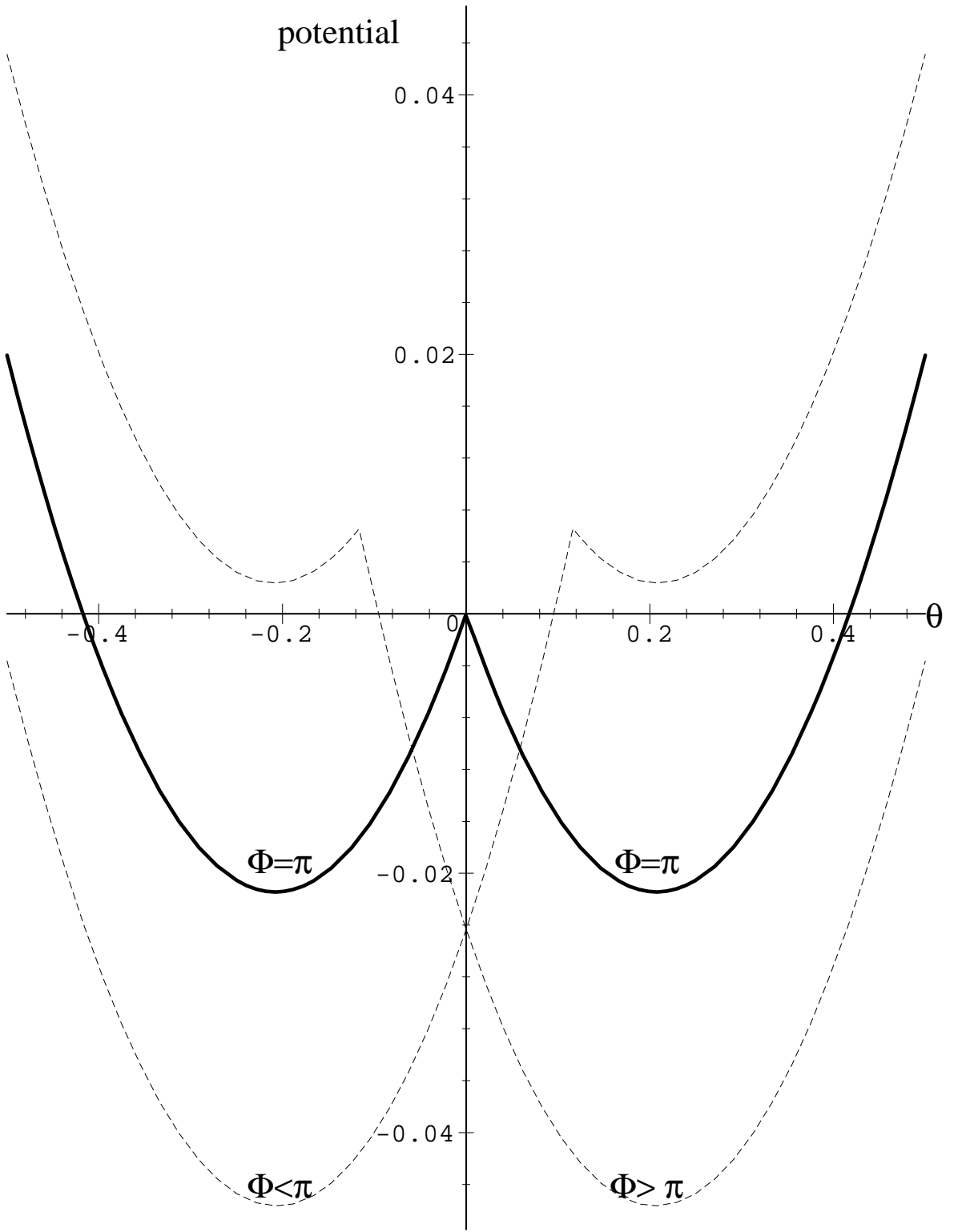
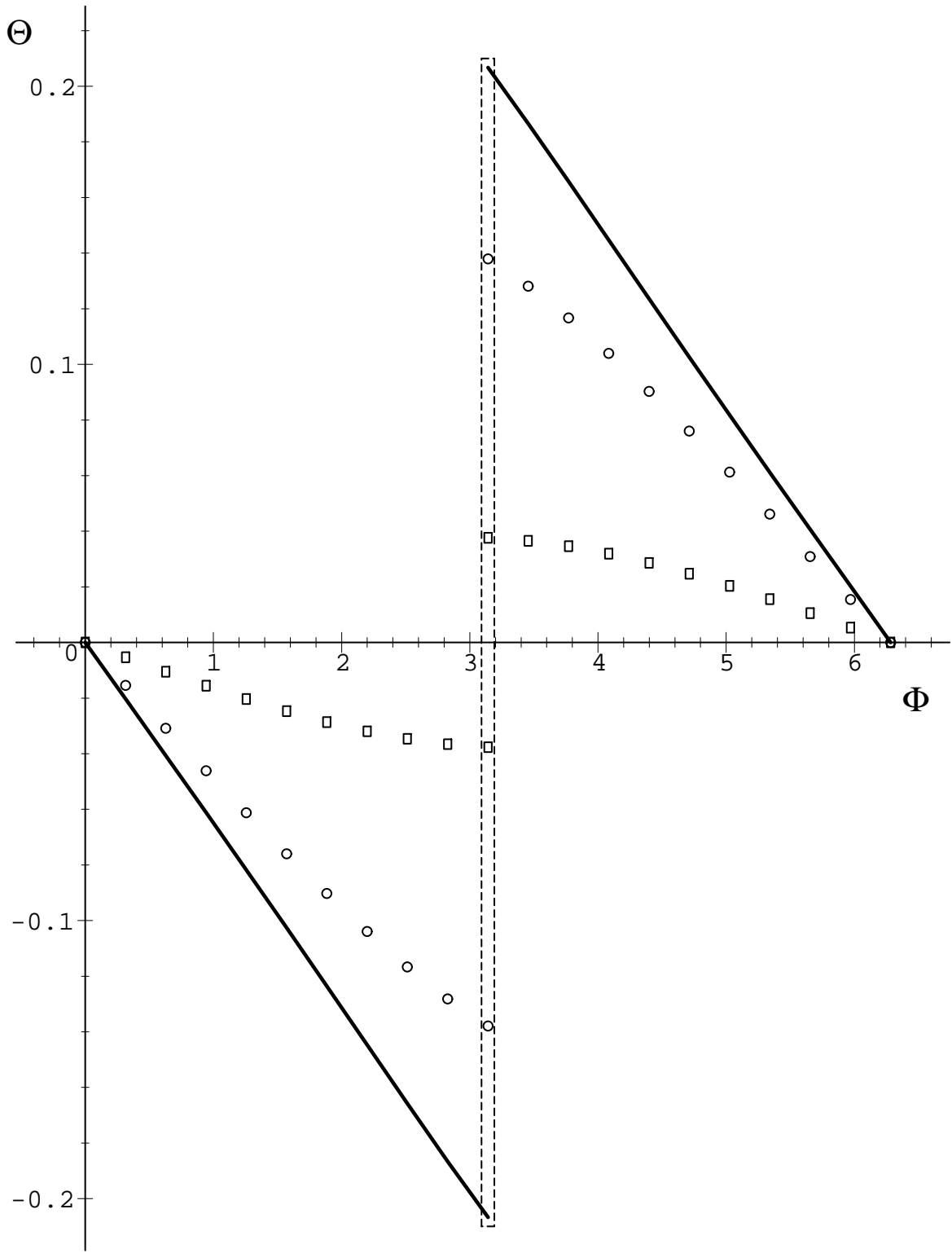
\end{document}